\documentstyle{article}
\begin{document}
\title{ Relaxation from color Compton Scattering of Plasmons in A Collisionless Quark-Gluon Plasma
\thanks{Supported by the National Natural Science Foundation of China Grant
No. 19805003
Email address  zhengxp@iopp.ccnu.edu.cn}}
\author{{Zheng Xiaoping$^{1 \hskip 2mm 2}$\ \ Chen Jisheng$^{2}$ Li Jiarong$^{2}$\ \ Liu
Lianggang$^1$\ \ Guo Shuohong$^{1}$}\\
{\small 1 Department of Physics,Zhongshan University,Guangzhou510275,P.R.China}\\
{\small 2 The Institute of Particle Physics, Huazhong Normal
University,Wuhan430079,P.R.China}}
\maketitle 

\begin{abstract}
\begin{minipage}{120mm}
Non-Abelian kinetic effects  
 are taken into account for the study of the relaxation of
collective motion in a quark-gluon plasma(QGP). An explicit Compton scattering 
is considered to calculate the  relaxation time.
 It is shown that the new non-Abelian relaxation has a physical
mechanism from the non-linear dynamics and the meaning of the
relaxation is discussed.
\end{minipage}
\end{abstract}
\vskip 0.9cm

The quark-gluon plasma(QGP) has been predicted to be produced in high energy
nuclear collisions. The properties of QGP  are of importance for
understanding experimental results\cite{r1}. Collective motion  plays an
importance role in a plasma. Remarkably, the collective behaviors of a QGP 
have  difference from a electromagnetic plasma due to color degrees
of freedom\cite{r2}. It means that some new transport problems 
 emerge, which is associated with color degrees of freedom\cite{r3,r4}.
 Obviously, the relaxation
processes from collisional term used to be easily realized since we are
familiar with the old knowledge.  The collisional terms of QGP thus were
given by Selikhov and Gyulassay\cite{r5} under the consideration of the quantum
fluctuations and the color relaxation from the color collisional term which is
only relevant to color degrees of freedom was studied as well
as momentum relaxation\cite{r6}. However, we can't help but ask such
questions:  Is there a relaxation related to both color and momentum space?
If there is, what is the relaxation? In this letter, we will try to solve
this
problem. Firstly, we need to review the development of QGP kinetic-theory
approach.
It is well-known that kinetic theory\cite{r7} is believed to
describe correctly the quark-gluon plasma(QGP) 
physics as well as temperature field\cite{r8} and kinetic-theory approach to
transport coefficients is of advantage of convenience. 
In fact,  more and more attention is recently concentrated on
the applications of kinetic theory to hot QCD\cite{r6,r9}. However,
the treatment of non-Abelian counterparts in kinetic equations had ever been a
difficult task\cite{r10,r11} for long time.
We just believed that the solution of the problem had
been made improvement until the non-Abelian mean-field dynamics\cite{r11} and 'double
perturbation' approach\cite{r12} were presented recently. 
These progresses are also
of importance for understanding color relaxation physics. We roughly give 
the analysis  before new physics are studied in detail: Apart from
collisional terms,  non-Abelian covariant derivative 
 enters  the formalism of the kinetic equations. This includes 
 color self-coupling contribution  in the covariant derivative. Here it is
clear that the  self-coupling term represents the collective behaviors
due to color degrees of freedom  instead of dynamic behaviors described in
collisional term in color space.  Then we think that a relaxation process 
from the non-linear dynamics shall
 also be produced  even if for a collisionless QGP. 

Now we start our studies from the kinetic equations of a collisionless
QGP\cite{r13},
\begin{equation}
p^\mu D_\mu Q_{\pm}({\bf p},x)\pm {g\over
2}p^\mu\partial^\nu_p\{F_{\mu\nu}(x),Q_\pm ({\bf p},x)\}=0,
\end{equation}
\begin{equation}
p^\mu {\tilde D}_\mu G({\bf p},x)+{g\over 2}p^\mu\partial^\nu_p\{{\tilde
F}_{\mu\nu}(x),G({\bf
p},x)\}=0,
\end{equation}
where the  letters with \~{} represent the corresponding operators in
adjoint representation of SU(3).  Here we consider 
the density fluctuations $\Delta Q_\pm, \Delta G$  deviating
from equilibrium distribution functions $Q^{(0)}_\pm$ and $G^{(0)}$ and
replace the induced field $A$ by $a$.  
 Assuming the fluctuations to be weak,
$p\sim gT, a_\mu\sim T, i\partial_\mu\sim gT$.
Thus we write the equations
for the fluctuations from 'double perturbation' approach\cite{r12}
\begin{eqnarray}
p^\mu\partial_\mu\Delta Q_\pm + ig\sum\limits_\lambda p^\mu[ a_\mu,
\Delta Q^{(\lambda)}_\pm]\pm gp^\mu f_{\mu\nu}\partial^\nu_p Q_\pm^{(0)}
=0
\end{eqnarray}
\begin{eqnarray}
p^\mu\partial_\mu \Delta G + ig\sum\limits_\lambda p^\mu[\tilde{ a}_\mu,
\Delta G^{(\lambda)}]+gp^\mu\tilde{f}_{\mu\nu}\partial^\nu_p G^{(0)}
=0
\end{eqnarray}
where $\lambda$ denotes the powers of induced field. In Ref\cite{r12}, 
we had showed the above
perturbation equations do not break non-Abelian gauge symmetry.
Marking the summation terms by $S_\pm$ and $\tilde {S}$, we give
\begin{eqnarray}
S_\pm=&\mp&\int{d^4k\over (2\pi)^4}\partial_p^\nu Q_\pm^{(0)}
\left ( ig\int {d^4k_1\over (2\pi)^4}{d^4k_2\over (2\pi)^4}\delta (k-k_1-k_2)
{1\over p\cdot k_2}[p\cdot a(k_1), p^\mu f_{\mu\nu}(k_2)]\right.\nonumber\\
&+&ig^2\int {d^4k_1\over (2\pi)^4}{d^4k_2\over (2\pi)^4}{d^4k_3\over (2\pi)^4}
\delta (k-k_1-k_2-k_3)
{1\over p\cdot (k_2+k_3)p\cdot k_2}[p\cdot a(k_1), [p\cdot a(k_2), p^\mu
f_{\mu\nu}(k_2)]]\nonumber\\
&+&\left.\cdots\cdots\right )
\end{eqnarray}
\begin{eqnarray}
\tilde {S}=&-&\int{d^4k\over (2\pi)^4}\partial_p^\nu G^{(0)}
\left ( ig\int {d^4k_1\over (2\pi)^4}{d^4k_2\over (2\pi)^4}\delta (k-k_1-k_2)
{1\over p\cdot k_2}[p\cdot \tilde {a}(k_1), p^\mu \tilde{f}_{\mu\nu}(k_2)]\right.\nonumber\\
&+&ig^2\int {d^4k_1\over (2\pi)^4}{d^4k_2\over (2\pi)^4}{d^4k_3\over (2\pi)^4}
\delta (k-k_1-k_2-k_3)
{1\over p\cdot (k_2+k_3)p\cdot k_2}[p\cdot\tilde{ a}(k_1), [p\cdot \tilde{a}(k_2), p^\mu
\tilde{f}_{\mu\nu}(k_2)]]\nonumber\\
&+&\left.\cdots\cdots\right )
\end{eqnarray}
We move the summation terms to the right-hand side of the
equations, a quasilinear formalism of the kinetic equations is obtained 
\begin{eqnarray}
p^\mu\partial_\mu\Delta Q_\pm \pm gp^\mu f_{\mu\nu}\partial^\nu_p Q_\pm^{(0)}
=-ig\sum\limits_\lambda p^\mu[ a_\mu,
\Delta Q^{(\lambda)}_\pm]
\end{eqnarray}
\begin{eqnarray}
p^\mu\partial_\mu \Delta G +gp^\mu\tilde{f}_{\mu\nu}\partial^\nu_p G^{(0)}
=-ig\sum\limits_\lambda p^\mu[\tilde{ a}_\mu,
\Delta G^{(\lambda)}]
\end{eqnarray}

Especially note that an illustration of the above equations is here
essential: In previous works, the nonlinear terms in the right-hand sides of
the equations were  removed because of the linearization treatment of
non-Abelian covariant derivatives in the kinetic equations\cite{r4,r6,r14},
so that only the relaxation from collisional terms were discussed. Now 
 we effectively regard the right-hand side terms 
 as 'collisional terms', but do not consider truly collisional terms. It is
the terms lost in past works that will give rise to a considerably large physical
effect, which  will be seen later.
In relaxation-time approach, we have identities from (5),(6),(7)and (8)
\begin{eqnarray}
p^\mu u_\mu\nu_\pm\Delta Q_\pm =-igS_\pm
\end{eqnarray}
\begin{eqnarray}
p^\mu u_\mu\nu_g\Delta G=-ig\tilde {S}
\end{eqnarray}
where $\nu$ is the equilibration rate parameter and $u_\mu$ is the
hydrodynamic velocity which describes the motion of the plasma as a whole.
After  the Fourier transformation, the equations (9)and(10) are respectively
multiplied by ${\bf v}\cdot {\bf a}(\omega, {\bf k'})$ and 
${\bf v}\cdot \tilde{\bf a}(\omega, {\bf k'})$ and then the mean values are
taken with respect to statistical ensemble and momentum  space. Thus one finds 
in the plasma rest frame:
\begin{eqnarray}
\nu_\pm p_0\int {{\rm d}^3p\over (2\pi)^3}\langle\Delta Q_\pm (k){\bf v\cdot
a}(k')\rangle=-g\int {{\rm d}^3p\over (2\pi)^3}{\rm Im}\langle S_\pm (k){\bf v\cdot
a}(k')\rangle,
\end{eqnarray}
\begin{eqnarray}
\nu_gp_0\int {{\rm d}^3 p\over (2\pi)^3}\langle \Delta G(k){\bf
v\cdot{\tilde a}}(k')\rangle=-g\int {{\rm d}^3p\over (2\pi)^3}{\rm
Im}\langle\tilde{S}(k){\bf v\cdot\tilde a}(k')\rangle,
\end{eqnarray}
where ${\bf v}={{\bf p}\over p_0}$ is the velocity of plasma particle.
One should note that for a baryonless plasma the numbers of quarks and antiquarks
are equal to each other and $\nu_+=\nu_-$ and a effective equilibration
parameter$\nu_{\rm eff}$ is defined via\cite{r14}
\begin{eqnarray}
\nu_{\rm eff}=\nu_+{N_f\over N_f+2N}+\nu_g{2N
\over N_f+2N}.
\end{eqnarray}
In general, it is sufficient that the first two terms in $S$ are remained, the
first term of which represents 3-wave correlator and be believed to
vanish\cite{r15}. So only the terms for $\lambda =2$ are  required to be
calculated here.  Then we
obtain the relaxation time formula as
{\small
\begin{eqnarray}
{1\over t_{pc}}&=&\nu_{\rm eff}\nonumber\\
&=&g^2N{
\int{\rm d}{\bf v}{{\rm d}k_1\over (2\pi)^4}\pi\delta[\omega_1-\omega-({\bf
k}_1-{\bf k})\cdot {\bf v}]
({\omega_1\over\omega_1-{\bf k}_1\cdot{\bf v}}-{\omega\over\omega-{\bf k\cdot v}})
({({\bf k}_1\cdot{\bf v})^2\over {\bf k}^2_1}\langle a_l^2(k_1)\rangle
 +{({\bf k}_1\times{\bf v})^2\over {\bf k}^2_1}\langle a_t^2(k_1)\rangle)
 \langle{\bf v\cdot a}(k)\rangle\over
 \int {\rm d}{\bf v}{\omega\over\omega-{\bf k\cdot v}}\langle{\bf v\cdot
a}(k)\rangle
 }
 \end{eqnarray}
}
 where $l$ and $t$ respectively denote the longitudinal and transverse
 components of the field.

Clearly, the delta-function 
$\delta[\omega_1-\omega-({\bf k}_1-{\bf k})\cdot {\bf v}]$ indicates the
a Compton scattering process of two collective modes(plasmons)in the phase
space with both color and momentum degrees of freedom.
To clarify the scattering process, the explanations follow: First, both color
and momentum exchanges
arise in the Compton scattering process; Second, the color exchange differs
from that without momentum; Finally, the Compton scattering is also not
usual Compton scattering in momentum space. Therefore, a non-Abelian Compton
scattering (or called color Compton scattering) mechanism shall be defined.
We also know that the scattering process is from the long-range
interactions and then
 $\omega > |{\bf k}|$ for any thermal collective mode from the
dispersion relation given by the past works\cite{r7,r13}.
 So we obtain in long-wavelength limit approximation
\begin{equation}
{1\over t_{\rm pc}}=0.236227g^2T.
\end{equation}

The  order in $g$ is  between the  momentum relaxation ${1\over t_{\rm p}}$
being of order of $g^4{\rm ln}g^{-2}$  and color relaxation ${1\over t_{\rm
c}}$  being of order of $g^2{\rm ln}g^{-1}$\cite{r4,r6}.To make $t_{\rm pc}$ clear,
 a analysis is necessary:
  
1. Color relaxation denoted by $t_{\rm c}$ is from static limit\cite{r6},
i.e., ${\bf v}=0$. It shall be a purely dynamic effect from collision
term  only considering color degrees of freedom\cite{r3}. 
While what we are studying is the non-Abelian Compton
scattering of two collective modes which is influenced by both color degrees
of freedom and plasma particle motions(${\bf v}\not=0$).  So the relaxation
process denoted by $t_{\rm pc}$ shall be a kinematic effect with color
dynamics. We can simply call it momentum-color-relaxation abbreviated to
pc-relaxation here.

2.Since the pc-relaxation is related to both color and motion, the relaxation
process will contributes to both the color diffusion and the momentum diffusion.
 It is know that ${1\over t_{\rm c}}$ was predicted to
vanishes in semiclassical approximation\cite{r4} and ${1\over t_{\rm
p}}$($\sim g^4T\log g^{-2}$)
for the perturbative QCD plasma\cite{r3} is of lower order than ${1\over t_{\rm
pc}}$($\sim g^2T$). Then pc-relaxation is of important interest in the semiclassical
limit domain.

3. The result in static limit can also be obtained from  the non-Abelian
mean-field dynamics based on kinetic-theory\cite{r11}. There it was believed
that the non-perturbative dynamics of soft gluons at leading logarithmic
order can be described\cite{r11, r16, r17}. It maybe implies that pc-relaxation instead
of color-relaxation controls the color diffusion coefficient in high
temperature  QGP.

4. The pc-relaxation process shall be produced due to the non-linear dynamics from
(3)-(6). We also find from (14) that two-mode scattering or
'two-quasiparticle collision' is the physical mechanism of the pc-relaxation.
 The effective collision,  in fact, is the long range interaction.

In conclusion, we find a kinetic effect on color  and momentum relaxations
from analysis of non-Abelian kinetic theory. After that, we study the
relaxation process and evaluate the relaxation time. We think what we get
in this letter is of benefit to deeply understanding and fully receiving
non-Abelian transport problem.


\begin{thebibliography}{99}
\bibitem{r1} E.V.Shuryak, Phys.Rep.61c,71(1980);D.J.Gross, R.D.Pisarski and
L.G.Yaff, Rev.Mod.Phys.53,43(1981);N.P.Landsman and Ch.G.van Weert,
Phys.Rep.145,141(1987);F.R.Brown et al.,
Phys.Lett.B251,181(1990);Phys.Rev.Lett.65,2491(1990);See reviews and
references in proceedings of Quark Matter91,edited by F.Plasil,
Nucl.Phys.A544,1c(1992).
\bibitem{r2} J.-P.Blaizot and E.Iancu, Phys.Rev.Lett.
70,3376(1993);72,3317(1994);Phys.Lett.B326,138(1994);Nucl.Phys.B421,
565(1994);434,663(1995).
\bibitem{r3} P.Danielewicz and M.Gyulassy,
Phys.Rev.D31,51(1985);H.Heiselberg, (1994).
\bibitem{r4} A.V.Selikhov and Gyulassy, Phys. Lett. B316,373(1993).
\bibitem{r5} A.V.Selikhov, Phys.Lett.B268,263(1991).
\bibitem{r6} A.V.Selikhov and Gyulassy, Phys. Rev.C49,1726(1994).
\bibitem{r7} U.Heinz,Phys.Rev.Lett.51,351(1983);56,
93(c)(1986);Ann.Phys.(N.Y.)161,48(1985);168,148(1986);\\
H.-Th.Elze, M.Gyulassy and D.Vasak,Phys.Lett.
B177,402(1986).
\bibitem{r8} D.J.Gross, R.D.Pisarski and L.G.Yaff, Rev.Mod.Phys.
53,43(1981);R.D.Pisarski, Phys. Rev.Lett.63,1129(1989);\\
E.Braaten and R.D.Pisarski, Nucl.Phys. B337,569(1990);J.Frenkel and
J.C.Taylor, Nucl.Phys. B334,199(1990).
\bibitem{r9} Zhang Xiaofei and Li Jiarong, Phys.
Rev.C52,964(1995);J.Phys.G21,1483(1995);Zheng Xiaoping and Li Jiarong,
Phys.Lett.B409,45(1997);Nucl.Phys.A639,705(1998);
 Yu.A.Markov and M.A.Markova, hep-ph/9902397,to published by
Transp.Theor.Stat.Phys.
\bibitem{r10} P.F.Kelly, Q.Liu,C C.Lucchesi and C.Manuel, Phys.Rev.Lett.
72,3461(1994); D50,4209(1994).
\bibitem{r11} D.F.Litim and C.Manuel, Phys.Rev.Lett. 82, 4991(1999);
hep-ph/9906210.
\bibitem{r12} Zheng Xiaoping and Li Jiarong, Submitted to Phys.Rev.Lett.
hep-ph/9911325.
\bibitem{r13} H.-Th. Elze and U.Heinz,Phys.Rep. 183,81(1989).
\bibitem{r14} S.Mr\'{o}wczy\'{n}ski, Phys.Rev. D39,1940(1989).
\bibitem{r15} V.N.Tsytovich, Theory of turbulent plasma,
Plenum,New.York,1977;
 A.G.Sitenko, Fluctuations and Non-linear Wave interactions in
plasma, Pergamon, Oxford, 1990.
\bibitem{r16} D.B\"{o}deker, Phys.Lett. B426,351(1998).
\bibitem{r17} J.-P. Blaizot and E.Iancu, hep-ph/9903389.
\end{thebibliography}
\end{document}